\documentclass[letterpaper, 10 pt, conference]{ieeeconf}  

\IEEEoverridecommandlockouts                             
\usepackage{amsmath,amsfonts}
\usepackage{algorithmic}
\usepackage{algorithm}
\usepackage{array}
\usepackage[caption=false,font=normalsize,labelfont=sf,textfont=sf]{subfig}
\usepackage{textcomp}
\usepackage{stfloats}
\usepackage{url}
\usepackage{verbatim}
\usepackage{graphicx}
\usepackage{cite}

\usepackage{orcidlink}
\usepackage{psfrag}
\usepackage{multirow}
 \usepackage{dsfont}
 \usepackage{pstricks}
 \usepackage{verbatim}
 \usepackage{graphicx}
 \usepackage{psfrag}
 \usepackage{fancybox}
 \usepackage{amsmath}
 \usepackage{amsfonts}
 \usepackage{amssymb}
 \usepackage{listings}
 \usepackage{enumerate}
 \usepackage{url}
 \usepackage{cancel}

 \psset{unit=1.0\unitlength}
 \SpecialCoor

 \newcounter{clcl}

\newcommand{\flowright}[1]{\begin{picture}(0,12)
   \put(0,-1){\makebox(0,0)[t]{$\rightarrow$}}
\end{picture}}

\newcommand{\flowleft}[1]{\begin{picture}(0,12)
   \put(0,-1){\makebox(0,0)[t]{$\leftarrow$}}
\end{picture}}

\newcommand{\ts}{\textstyle}

\def\M{{\bf M}}

 \input{pst-plot}
 \input{pst-node}
 \input{pst-coil}

 \makeatletter
 \newcommand\notsotiny{\@setfontsize\notsotiny\@vipt\@viipt}
 \makeatother

 \newcommand{\ShortFig}{_Short}

 \renewcommand{\M}{M}
 \renewcommand{\gamma}{M_r}

 \newcommand{\db}{\mbox{ \rm db}}
 \newcommand{\reldeg}{\tilde{r}}

 \definecolor{verylightgrey}{rgb}{0.68, 0.68, 0.68}
 \definecolor{verde}{rgb}{0.0, 0.5, 0.0}
 \definecolor{arancione}{rgb}{0, 1.0, 1.0}

\renewcommand{\keywords}[1]{\textbf{\textit{Index terms---}} #1}

  \newtheorem{Prop}{\bf Property}
  \newtheorem{Defini}{\bf Definition}
 \newtheorem{Note}{\bf Note}

 \psfrag{Diagramma asintotico dei moduli}[b][b][0.7]{}
 \psfrag{Diagramma a gradoni delle fasi}[b][b][0.7]{}
 \psfrag{Diagramma dei moduli}[b][b][0.7]{Magnitude Plot}
 \psfrag{Diagramma delle fasi}[b][b][0.7]{Phase Plot}

\title{\LARGE \bf
The Construction of Asymptotic Bode Plots: A New Direct Method
}

\author{Davide Tebaldi\orcidlink{0000-0003-1432-0489} and Roberto Zanasi\orcidlink{0000-0001-5507-825X}
\thanks{The work was partly supported by the University of Modena and Reggio Emilia
through the action FARD (Finanziamento Ateneo Ricerca Dipartimentale) 2023/2024, and funded under the National Recovery and Resilience Plan (NRRP), Mission 04 Component 2 Investment 1.5 - NextGenerationEU, Call for tender n. 3277 dated 30/12/2021
Award Number:  0001052 dated 23/06/2022.}
\thanks{The authors are
with the Department of Engineering ``Enzo Ferrari'', University of
Modena and Reggio Emilia, Modena, Via Pietro Vivarelli 10, 
41125 Modena, Italy. E-mails: \{davide.tebaldi, roberto.zanasi\}@unimore.it. }
\thanks{This work has been submitted to the IEEE for possible publication. Copyright may be transferred without notice, after which this version may no longer be accessible.}
}

\begin{document}

\maketitle
\thispagestyle{empty}
\pagestyle{empty}

\begin{abstract}

Bode plots represent an essential tool in control and systems engineering. In order to perform an initial qualitative analysis of the considered systems, the construction of asymptotic Bode plots is often sufficient. The standard methods for constructing asymptotic Bode plots are characterized by the same drawbacks: they
are not systematic, may be not precise and time-consuming. This
is because they require the detailed analysis of the different
factors composing the considered transfer function, meaning that
more and more intermediate steps are required as the number
of factors increases. In this paper, a new method for the construction of asymptotic Bode plots is proposed, which is based on the systematic
calculations of the so-called generalized approximating functions
and on the use of well defined properties. The proposed method
is referred to as a direct method since it allows to directly draw
the asymptotic Bode magnitude and phase plots of the complete
transfer function without requiring the detailed analysis nor the
plots construction of each factor.
This latter feature also makes the proposed direct method more systematic, potentially more precise and less time-consuming compared to standard methods, especially when dealing with a large number of factors in the transfer function.
The comparison of the proposed direct method with the standard approaches is performed, in order to examine the benefits offered by the direct method.

\end{abstract}


\section{INTRODUCTION}

Bode plots, Nyquist plots and Nichols plots~\cite{Nuovo_6} are different graphical representations of the frequency response function, which exhibit different characteristics with respect to each other while carrying the same informative content.

Nyquist and Nichols plots are composed of a single diagram, sketching imaginary versus real part and modulus versus phase in the case of Nyquist and Nichols plots, respectively. Nichols plots employ decibels, a logaritmic unit of measurement, for the magnitude representation, and therefore allow for an effective representation of both large and low magnitude values. Despite providing a less detailed representation of the frequency response function,
Nyquist plots find applications in control theory such as the Nyquist stability criterion~\cite{Nyquist_9}, the passivity analysis of physical systems, and the design of lead/lag networks~\cite{Nyquist_8}.
  On the other hand, Bode plots also employ decibels for the magnitude representation, and therefore share a common advantage with Nichols plots. However, Bode plots are composed of two separated diagrams for the phase and magnitude representation, respectively, as a function of the angular frequency, and therefore allow for a more detailed representation of the frequency response function compared to Nyquist and Nichols plots.

Bode plots were invented in 1938 by Hendrik Wade Bode \cite{Nuovo_8}, and are widely used nowadays for the analysis of linear systems~\cite{LTI_Syst_1,LTI_Syst_1_biss,LTI_Syst_1_biss_biss}. Their uses in control theory include the stability analysis of closed-loop systems, through the Bode stability criterion~\cite{Nuovo_1} and through the reading of the phase and gain margins of the system~\cite{LTI_Syst_2}.
Bode plots are also used to evaluate the dynamic performances of open and closed-loop systems by reading the bandwidth and, consequently, the rising time~\cite{LTI_Syst_4}, as well as to design different types of compensators \cite{LTI_Syst_3,LTI_Syst_3_bisss} that can be used for different applications~\cite{LTI_Syst_5} including DC motor control \cite{LTI_Syst_5_biss}.

The usefulness of Bode plots extends to other engineering and non-engineering fields as well. In~\cite{Nuovo_2}, Bode plots are used to perform the graphical analysis of electrochemical impedance spectroscopy data, while in~\cite{Nuovo_3} and \cite{Nuovo_3_biss} they are used to
evaluate the corrosion in micro/nanocapsulated polymeric coatings and to perform the frequency response of a simplified tank model, respectively. 

Whenever a detailed and punctual analysis of the considered system is desired, the Bode magnitude and phase plots are typically constructed by making use of computer aided control system design software, such as the Control System Toolbox available in the MATLAB/Simulink environment~\cite{CST_Ref}, LabVIEW~\cite{LabVIEW_Octave_Ref}, and others.
However, precision is sometimes not the primary goal for a first qualitative system analysis, while speed and qualitative
behavior may be the objectives instead. In these cases,
asymptotic Bode plots are preferred~\cite{As_Bode_1}. The manual construction of asymptotic Bode plots is indeed still a necessary knowledge which every control engineer should have for different purposes, including the capability of performing an initial qualitative analysis of the system under consideration.

The standard methods for the construction of asymptotic Bode plots can be
classified into two approaches. The first approach consists into the identification of the different factors composing the considered transfer function~\cite{Nuovo_6_bis}. Once the factors have been identified, the asymptotic Bode magnitude and phase plots of each factor need to be plotted. The asymptotic Bode magnitude and phase plots of the complete transfer function can finally be obtained by
adding together the asymptotic plots of each factor.
As far as the second approach is concerned, the asymptotic magnitude plot is obtained by first constructing the low-frequency magnitude asymptote, and then introducing the changes in slope given by the different factors~\cite{Nuovo_6}. The asymptotic phase plot is obtained by first computing the low-frequency phase, and then introducing the changes in phase given by the different factors~\cite{Nuovo_6}.
The standard methods for constructing asymptotic Bode plots are characterized by the same drawbacks: they are not systematic, may be not precise and time-consuming.
This is because they require the detailed analysis of the different factors composing the considered transfer function, meaning that more and more intermediate steps are required as the number of factors increases. In this paper, a
new method for the construction of asymptotic Bode plots is proposed, which is based on the systematic calculations of the so-called {\it generalized} approximating functions and on the use of well defined properties. The proposed method is referred to as a {\it direct} method since it allows to directly draw the asymptotic Bode magnitude and phase plots of the complete transfer function without requiring the detailed analysis nor the plots construction of each factor.

This paper is structured as follows.
The contextualization and definitions needed for the proposed method are first given in Sec.~\ref{appr_sect}, followed by the presentation of the proposed direct method for the construction of asymptotic Bode plots in Sec.~\ref{Asymptotic_Bode_Plots}.
In order to make the proposed method easily understandable and accessible,
it is applied to three different case studies in Sec.~\ref{case_studies_sect} and Sec.~\ref{comparison_sect}. Furthermore, Sec.~\ref{comparison_sect} also addresses the comparison of the proposed direct method with the standard approaches, showing the advantages provided by the proposed direct method. The conclusions of this work are given in Sec.~\ref{Conclusion_sect}.

\section{Context and Definitions}\label{appr_sect}
Reference is made to the following transfer
function $G(s)$:
\begin{equation}\label{Gs_Fact}
  G(s)
  =
  \frac{K}{s^h}\frac{ \prod_{i=1}^m p_i^{r_i}(s)}{\prod_{i=m+1}^{\bar{r}} p_i^{r_i}(s)},
\end{equation}
where $h$
is the number of poles at the origin of function $G(s)$. The polynomial terms $p_i(s)$ have the following structure:
\[
  p_i(s) \!=\!
  \left\{
  \begin{array}{@{\!}l@{\;\;}l}
  a_{i 1} s\!+\!a_{i 0}   & \mbox{if } p_i(s) \mbox{ is a first-order term},  \\[1mm]
  a_{i 2} s^2\!+\!a_{i 1} s\!+\!a_{i 0}  & \mbox{if } p_i(s) \mbox{ is a second-order term},
  \end{array}
  \right.
 \]
 where
  $a_{i1}\!\neq \!0$, $a_{i0}\!\neq \!0$,
  $a_{i2}\!\neq \!0$,  $a_{i0}\!\neq \!0$ and the roots $\lambda_{1,2}$ of the second-order terms $p_i(s)$ are supposed to be complex conjugate numbers. The parameter $r_i$ in \eqref{Gs_Fact} is the degree of multiplicity of the $i$-th polynomial term $p_i(s)$.
 The total number of polynomial terms $p_i(s)$ of function $G(s)$ is $\bar{r}$, out of which the first $m$ terms belong to the numerator while the last $\bar{r}-m$ terms belong to the denominator.
 The numbers $n_z$ and $n_p$ of zeros and poles of function $G(s)$ are:
$ n_z=\sum_{i=1}^m r_i$ and $n_p=h+\sum_{i=m+1}^{\bar{r}} r_i$.

\begin{Defini}\label{critical_freq_defini_many}
The polynomial terms $p_i(s)$ are characterized by the following parameters:
 \\
1) the {\it order} $n^o_i$ of term $p_i(s)$:
  \begin{equation}\label{order}
  n^o_i=
  \left\{
  \begin{array}{@{\;}l@{\;\;}l}
  1   & \mbox{ if } p_i(s) \mbox{ is a first-order term},  \\[0mm]
  2  & \mbox{ if } p_i(s) \mbox{ is a second-order term}.
  \end{array}
  \right.
  \end{equation}
2) the {\it main root} $\lambda_i$ of term $p_i(s)$:
  \begin{equation}\label{main_root}
\lambda_i=
  \left\{
  \begin{array}{@{\;}l@{\;\;}l}
  \frac{a_{i 0}}{a_{i 1}}   & \mbox{ if } n^o_i=1,  \\[0mm]
  \frac{-a_{i 1}+j\sqrt{a_{i 1}^2-4 a_{i 2} a_{i 0}}}{2 a_{i 2}}  & \mbox{ if }  n^o_i=2.
  \end{array}
  \right.  \end{equation}
For second-order terms, i.e. $n^o_i=2$, the root $\lambda_i$ with positive imaginary part is considered.
\\[1mm]
3) the {\it zero-pole sign} $S_i^{zp}$ of term $p_i(s)$:
  \begin{equation}\label{pz_sign}
S_i^{zp}=
  \left\{
  \begin{array}{@{\;}l@{\;\;}l}
  1   & \mbox{ if root $\lambda_i$ is a zero: } i\leq m,
  \\[0mm]
-1   & \mbox{ if root $\lambda_i$ is a pole: } i> m.
  \end{array}
  \right.
  \end{equation}
4) the {\it stability sign} $S_i^{st}$ of term $p_i(s)$:
  \begin{equation}\label{st_sign}
S_i^{st}=
  \left\{
  \begin{array}{@{\;}l@{\;\;}l}
   1   & \mbox{ if root $\lambda_i$ is stable: $\mbox{real}(\lambda_i)\leq 0$,}
  \\[0mm]
  -1   & \mbox{ if root $\lambda_i$ is unstable: $\mbox{real}(\lambda_i)> 0$.}
  \end{array}
  \right.
  \end{equation}
5) the {\it critical frequency} $\omega_{ci}$ of term $p_i(s)$:
  \begin{equation}\label{lambda_i}
  \omega_{ci} = 
  \left\{
  \begin{array}{l@{\;\;\;\;\mbox{if}\;}l}
  \ts  \left|\frac{a_{i 0}}{a_{i 1}}\right|  &
  n^o_i=1,  \\[1mm]
  \ts  \sqrt{\frac{a_{i 0}}{a_{i 2}}}  &  n^o_i=2.  \\[1mm]
  \end{array}
  \right.
  \end{equation}
 6) the {\it highest order term} $a_{i}^*(s)$ of term $p_i(s)$:
  \begin{equation}\label{highest_order_coefficient}
  a_i^*(s) =
  \left\{
  \begin{array}{l@{\;\;\;\;\mbox{if}\;\;\;\;}l}
  a_{i 1}\,s  &  n^o_i=1,  \\[0mm]
  a_{i 2}\,s^2  &  n^o_i=2.
  \end{array}
  \right.
  \end{equation}
\end{Defini}

\subsection{Low and high frequency behaviors of   $G(s)$}

The low frequency behavior of  function $G(s)$
is described  by the following approximating function $G_0(s)$:
\begin{equation}\label{G0s_eq}
G_0(s)=\left.G(s)\right|_{s \simeq 0^+}=\frac{K_0}{{s}^h},
\end{equation}
where $h$ is the number of poles at the origin of function $G(s)$ and  $K_0$ is a constant term that results from \eqref{Gs_Fact} when $s \simeq 0^+$:
\[
K_0=
 K\,\frac{ \prod_{i=1}^m a_{i0}^{r_i}}{\prod_{i=m+1}^{\bar{r}} a_{i0}^{r_i}}.
\]
%
The high frequency behavior of function $G(s)$ is described  by the following approximating function $G_\infty(s)$:
\begin{equation}\label{Ginfs_eq}
G_\infty(s)= \left.G(s)\right|_{s \simeq \infty}=\frac{K_\infty}{{s}^{\reldeg}},
\end{equation}
where $\reldeg=n_p-n_z$ is the relative degree of function $G(s)$, and $K_\infty$ is a constant term that results from \eqref{Gs_Fact} when $s\!\simeq\! \infty$:
\[
K_\infty=
 K\,\frac{ \prod_{i=1}^m (a_{i}^*)^{r_i}}{\prod_{i=m+1}^{\bar{r}} (a_{i}^*)^{r_i}},
\]
where $a_{i}^*$ is the coefficient of the  highest order term $a_i^*(s)$  defined in \eqref{highest_order_coefficient}.

\section{Direct Construction of the Asymptotic Bode Plots}\label{Asymptotic_Bode_Plots}

\begin{Defini}\label{Omega_c}
  Let $\Omega_c$ denote the set of all the distinct critical frequencies $\omega_{ci}$ properly ordered in increasing order:
  \begin{equation}\label{omega_c_defini}
      \Omega_c=\mbox{unique}(\mbox{sort}(\{\omega_{c1},\,\ldots,\,\omega_{c\bar{r}}\}))=\{\omega_1,\,\ldots,\,\omega_{r}\},
  \end{equation}
  where $r\!\leq\!\bar{r}$.
  Let $\omega_{k}\!\in\!\Omega_c$ denote the $k$-th element of
  set $\Omega_c$.
\end{Defini}
\vspace{1mm}

\begin{Defini}\label{Sets_Ik}
Let $\mathcal{I}_k$
denote the following set:
  \begin{equation}\label{Ik}
 \mathcal{I}_k =\{i: 1\leq i \leq r\wedge\omega_{ci}=\omega_{k}\},
  \end{equation}
that is, the set of all the indices $i$ of the  polynomial terms $p_i(s)$ whose critical frequency $\omega_{ci}$ is equal to $\omega_k$.
\end{Defini}
\vspace{1mm}

\begin{Prop}\label{critical_freq_defini}
With reference to function $G(s)$ in (\ref{Gs_Fact}),
the $k$-th approximating function $G_k(s)$ is defined as follows:
  \begin{equation} \label{G_k_s}
  G_k(s)=\frac{K}{s^h} \frac{\prod_{i=1}^m q_i^{r_i}(s)}{\prod_{i=m+1}^{\bar r} q_i^{r_i}(s)}
  =\frac{K_k}{s^{t_k}},
 \end{equation}
where the polynomial terms $q_i(s)$ have the following structure:
  \begin{equation} \label{G_k_s_bis}
  q_i(s) =
  \left\{
  \begin{array}{l@{\;\;\;\;\mbox{if}\;\;\;\;}l}
  a_{i 0}  &  \omega_{ci}> \omega_{k}, \\[0mm]
  a_i^*(s)  &  \omega_{ci}\leq \omega_{k}.
  \end{array}
  \right.
 \end{equation}
\end{Prop}
 \vspace{2mm}
The constant $K_k$ in \eqref{G_k_s} is given by the products and ratios of constant $K$ in \eqref{G_k_s} and of coefficients $a_{i 0}$ and $a_i^*$ in \eqref{G_k_s_bis}. Parameter $t_k$ in \eqref{G_k_s} is the relative degree of function $G_k(s)$ and can be obtained
using the following recursive formula:
  \begin{equation} \label{t_k_rec}
 t_k
 = t_{k-1} - \sum_{i\in\mathcal{I}_k} S_i^{zp} n^o_i,
 \end{equation}
with initial condition $t_0\!=\!h$ and using the set  $\mathcal{I}_k$
in \eqref{Ik}.

For detailed examples of application of Property~\ref{critical_freq_defini}, the readers are referred to Sec.~\ref{first_case_study_sect}, Sec.~\ref{second_case_study_sect} and Sec.~\ref{Direct_Method_sect}.

\begin{Note}\label{G0_Goo_note}
The approximating functions $G_k(s)$ in \eqref{G_k_s} represent a generalization of the approximating functions $G_0(s)$ and $G_\infty(s)$ in \eqref{G0s_eq} and \eqref{Ginfs_eq}, respectively.
If $\omega_{k}=\omega_{0}=0 \notin \Omega_c$ in
\eqref{G_k_s}
is considered, the approximating function $G_0(s)$ in \eqref{G0s_eq} is obtained, which is valid in the frequency range $\omega \in [0,\,\omega_{1}]$, with $\omega_{1} \in \Omega_c$ in \eqref{omega_c_defini}.
If $\omega_{k}=\omega_{r} \in \Omega_c$ in
\eqref{G_k_s}
 is considered, the approximating function $G_r(s)=G_\infty(s)$ in \eqref{Ginfs_eq} is obtained, which is valid in the frequency range $\omega \in [\omega_{r},\,\infty]$.
\end{Note}

%
%
\subsection{Asymptotic
Magnitude Plot}\label{Bode_Magnitude_Plot}
\begin{figure}[t]
\centering
 \includegraphics[clip,width=0.95\columnwidth]{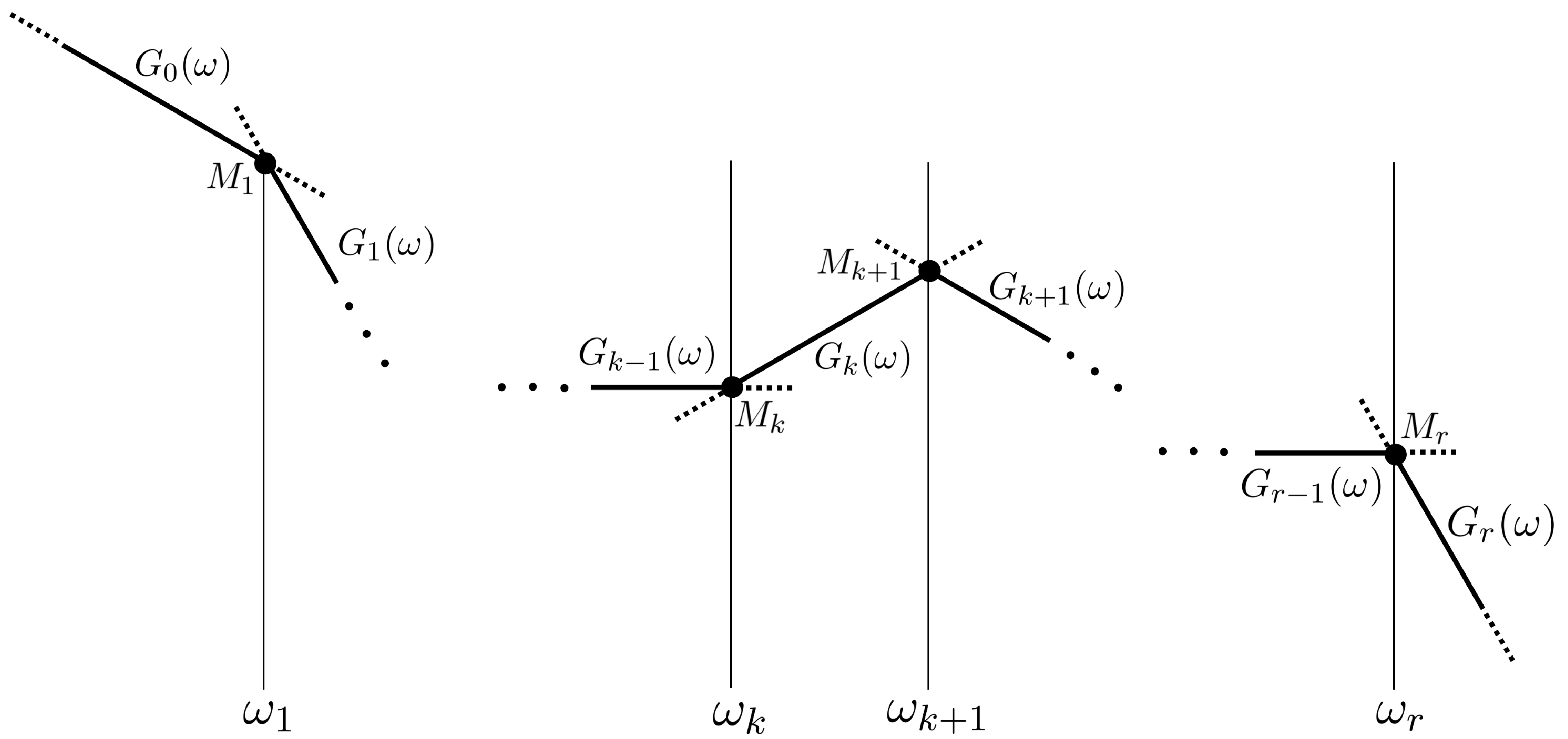}
    \vspace{-2mm}
    \caption{Graphical representation of the asymptotic Bode magnitude plot of function $G(s)$ in \eqref{Gs_Fact}. The latter has a piecewise behavior composed of a sequence of $r-1$ inner linear segments $G_k(\omega)=|G_k(s)|_{s=j\omega}$ for $\omega \in \mathcal{B}^\omega_{k}$ in \eqref{B_omega_k} computed as in Property~\ref{bode_magn_prop}, where $r=\mbox{dim}(\Omega_c)$ in \eqref{omega_c_defini}, and $2$ external linear segments.
}\label{bode_amplitude_schematic}
\vspace{-2mm}
\end{figure}
 The  asymptotic Bode magnitude plot  of a
 function $G(s)$ as in \eqref{Gs_Fact} has a piecewise behavior composed of a sequence of $r-1$ inner linear segments and $2$ external linear segments, as shown in Figure~\ref{bode_amplitude_schematic}.  The asymptotic Bode magnitude plot changes slope in correspondence of points $P_k=(\omega_k,\; M_k)$, where $M_k$  is the $k$-th  {\it critical gain}. The latter can be computed in two different ways using the following Property~\ref{critical_gain_defini} and Property~\ref{gains_formula}.
\begin{Prop}\label{critical_gain_defini}
For $k \in \{1,\,\ldots,\,r\}$, the  $k$-th critical gain $M_k$ can be computed as follows:
\begin{equation}\label{beta_K_example}
  \M_k
  =\left| G_{k-1}(s)\right|_{s=j\omega_{k}}
  =\left| G_{k}(s)\right|_{s=j\omega_{k}},
\end{equation}
where $G_{k-1}(s)$ and  $G_{k}(s)$ are the $(k-1)$-th and the  $k$-th approximating functions defined in \eqref{G_k_s}.
\end{Prop}
\vspace{1mm}

  \begin{Prop}\label{gains_formula}
The critical gains $\M_k$, for $k\in\{1,\,2,\,\ldots,\,r-1\}$,
can also be obtained by using the following recursive formula:
\begin{equation}\label{beta_K_example_bis}
 \M_{k+1} = \M_{k}\left(\frac{\omega_{k}}{\omega_{k+1}}\right)^{t_k\,},
\end{equation}
 where $t_k$ defined in \eqref{t_k_rec} is the relative degree of  function $G_k(s)$, and using the  initial condition: $\M_1 =\left| G_0(s)\right|_{s=j\omega_{1}}$.
  \end{Prop}
\vspace{1mm}

For detailed examples of application of Property~\ref{critical_gain_defini} and Property~\ref{gains_formula}, the readers are referred to Sec.~\ref{first_case_study_sect}, Sec.~\ref{second_case_study_sect} and Sec.~\ref{Direct_Method_sect}.

\begin{Defini}\label{critical_gain_defini_1}
Let $\mathcal{B}^\omega_{k}$ denote the following frequency band: 
\begin{equation}\label{B_omega_k}
\mathcal{B}^\omega_{k} = \{\omega : \omega\in[ \omega_k,\;\omega_{k+1}]\,\},
\end{equation}
for $k \in \{0,\,\ldots,\,r\}$ and where $\omega_0=0$ and  $\omega_{r+1}=\infty$. Additionally, note that the functions $G_k(\omega)$ present in  Figure~\ref{bode_amplitude_schematic} are defined as follows:
$G_k(\omega)=|G_k(s)|_{s=j\omega}$ for $\omega \in \mathcal{B}^\omega_{k}$.

\end{Defini}
\vspace{1mm}

  \begin{Prop}\label{bode_magn_prop}
 The $k$-th inner segment $G_k(\omega)$ defined in the frequency band $\mathcal{B}^\omega_{k}$ can also be computed as follows:
 \[
 G_k(\omega)=
 M_k + \frac{M_{k+1}-M_{k}}{\bar{\omega}_{k+1}-\bar{\omega}_{k}}(\bar{\omega}-\bar{\omega}_{k}),
 \]
where $\bar{\omega}=\log_{10}(\omega)$,  $\bar{\omega}_k=\log_{10}(\omega_k)$ and $k \in \{1,\,\ldots,\,r-1\}$. The first external segment $G_0(\omega)$ is constructed starting from the point $P_1=(\omega_1,\,M_1)$ and drawing a segment having a slope $-h$  congruent with the approximating function $G_0(s)$ defined in \eqref{G0s_eq}. The second external segment $G_r(\omega)$ is constructed starting from the point $P_r=(\omega_r,\,M_r)$ and drawing a segment having a slope $-\reldeg$ congruent with  the approximating function $G_r(s)=G_\infty(s)$ given in \eqref{Ginfs_eq}.
  \end{Prop}
\vspace{1mm}

Property~\ref{bode_magn_prop} is employed in
Sec.~\ref{first_case_study_sect}, Sec.~\ref{second_case_study_sect} and Sec.~\ref{Direct_Method_sect} for the construction of the asymptotic Bode magnitude plots.

\subsection{
Stepwise Phase Plot}\label{Bode_Phase_Plot}

\begin{figure}[t]
\centering
 \includegraphics[clip,width=0.95\columnwidth]{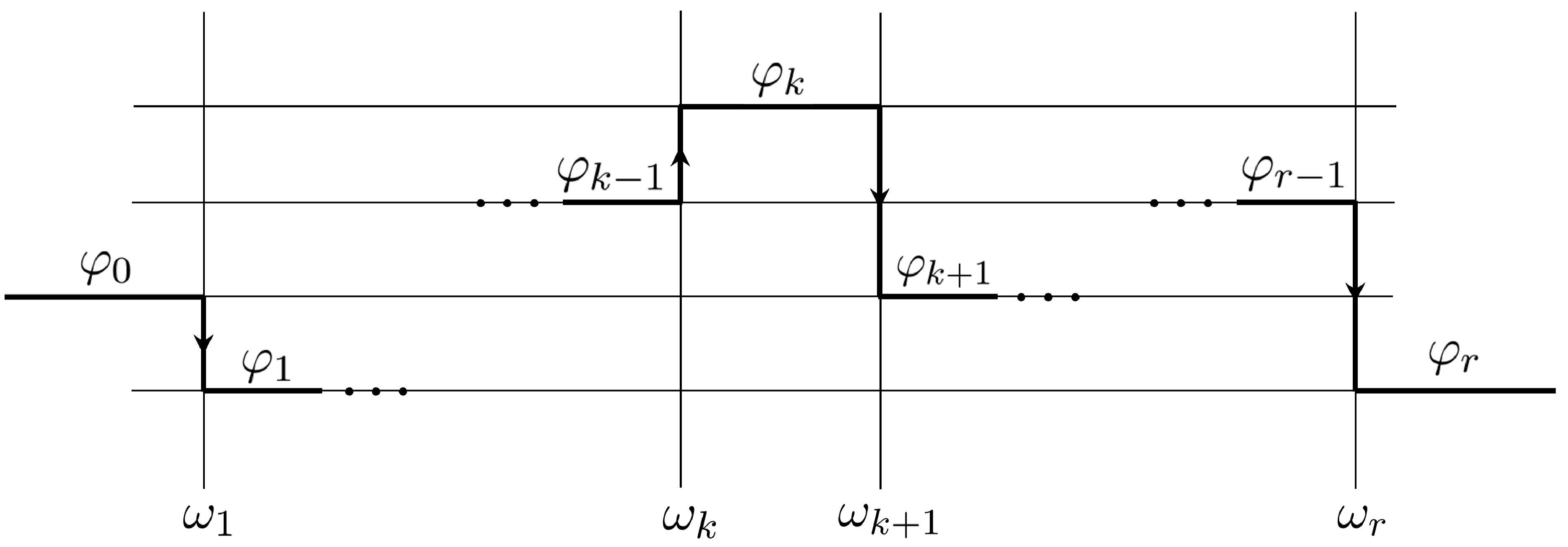}
    \vspace{-2mm}
    \caption{Graphical representation of the stepwise Bode phase plot of function $G(s)$. The latter has a piecewise behavior composed of a sequence of $r+1$ horizontal segments $\varphi_k(\omega)$ computed as in Property~\ref{critical_gain_defini_prop}, where $r=\mbox{dim}(\Omega_c)$ in \eqref{omega_c_defini}, and of $r$ vertical segments obtained by connecting the endpoints of the horizontal segments $\varphi_k(\omega)$.
    }\label{bode_phase_schematic}
    \vspace{-2mm}
\end{figure}
 The stepwise Bode phase plot  of a
 function $G(s)$ as in \eqref{Gs_Fact} has a piecewise behavior composed of a sequence of $2r+1$ linear segments, as shown in Figure~\ref{bode_phase_schematic}.

\begin{Prop}\label{critical_gain_defini_prop}
The $r+1$ horizontal segments $\varphi_k(\omega)$  of the  stepwise Bode phase plot defined within the frequency band  $\mathcal{B}^\omega_{k}$ can be computed
 by using the following recursive formula:
\begin{equation}\label{varphi_k}
\varphi_{k} = \varphi_{k-1} +  \sum_{i\in\mathcal{I}_k}\Delta\varphi_i,
\end{equation}
  where $\Delta\varphi_i$ is the total phase shift generated by all the polynomial terms $p_i(s)$, for $i\in\mathcal{I}_k$ in \eqref{Ik},
  in the vicinity of the critical frequency $\omega_{ci}=\omega_k$:
\[
\Delta\varphi_i  = r_i S_i^{zp} S_i^{st} n^o_i\, \frac{\pi}{2},
\]
 for $k \in \{1,\,2,\,\ldots,\,r\}$  and using the initial condition $\varphi_0  = \arg(G_{0}(j\omega))$.  When $\omega = \omega_k$,  the remaining $r$ vertical segments are obtained by connecting the endpoints of the segments $\varphi_k(\omega)$ in \eqref{varphi_k}, as shown in Figure~\ref{bode_phase_schematic}.
\end{Prop}
\vspace{1mm}

Property~\ref{critical_gain_defini_prop} is employed in Sec.~\ref{first_case_study_sect}, Sec.~\ref{second_case_study_sect} and Sec.~\ref{Direct_Method_sect} for the construction of the stepwise Bode phase plots.

\section{Case Studies}\label{case_studies_sect}

\subsection{First Case Study}\label{first_case_study_sect}
Reference is made to the following transfer function:
\begin{equation}\label{Gs_example}
 G(s)=\frac{60\,(s^2+0.8\,s+4)}{s(s-30)(\frac{s}{200}+1)^2}.
\end{equation}
 The polynomial terms of function $G(s)$ are: $p_1(s)\!=\!(s^2\!+\!0.8\,s\!+\!4)$,  $p_2(s)\!=\!(s\!-\!30)$ and $p_3(s)\!=\!(\frac{s}{200}\!+\!1)$.
  The set $\Omega_c$ of critical frequencies defined in \eqref{omega_c_defini} is  $\Omega_c=\{2,\;30,\;200\}$.
By applying \eqref{G_k_s}, the following approximating functions $G_k(s)$ can be extracted from function $G(s)$ in \eqref{Gs_example}:
\begin{equation}\label{Approx_Gs_example}
 \begin{array}{@{}r@{\,}c@{\,}l@{\;\;\mbox{for}\;\;}c@{}}
 G_0(s)
 &=&
 \frac{60\,({\color{verylightgrey}\cancel{s^2}+\cancel{0.8\,s}}+4)}{s({\color{verylightgrey}\cancel{s}}-30)({\color{verylightgrey}\cancel{\frac{s}{200}}}+1)^2}
 =-\frac{8}{s}
 &
 \omega \leq 2,
\\[2mm]
 G_1(s)
 &=&
 \frac{60\,(s^2{+\color{verylightgrey}\cancel{0.8\,s}+\cancel{4}})}{s({\color{verylightgrey}\cancel{s}}-30)({\color{verylightgrey}\cancel{\frac{s}{200}}}+1)^2}
 =-2\,s
  &
 2\leq \omega \leq 30,
\\[2mm]
 G_2(s)
 &=&
 \frac{60\,(s^2{\color{verylightgrey}+\cancel{0.8\,s}+\cancel{4}})}{s(s
 {\color{verylightgrey}-\cancel{30}})({\color{verylightgrey}\cancel{\frac{s}{200}}}+1)^2}
 =60
  &
 30\leq \omega \leq 200,
\\[2    mm]
  G_3(s)
 &=&
 \frac{60\,(s^2{\color{verylightgrey}+\cancel{0.8\,s}+\cancel{4}})}{s(s
 {\color{verylightgrey}-\cancel{30}})(\frac{s}{200}{\color{verylightgrey}+\cancel{1}})^2}
 =\frac{2.4\,10^6}{s^2}
  &
 \omega \geq 200.
 \end{array}
\end{equation}

\begin{figure}[t]
  \psfrag{Phioo}[l][l][0.6]{$\varphi_\infty$}
 \psfrag{Phi0}[lb][lb][0.6]{$\varphi_0$}
 \psfrag{G0s}[l][l][0.6]{$G_0(s)$}
 \psfrag{Goos}[l][l][0.6]{$G_\infty(s)$}
 \psfrag{be}[lb][lb][0.5]{$M_1$}
 \psfrag{ga}[l][l][0.5]{$M_3$}
\psfrag{M0}[t][t][0.7]{}
\psfrag{M1}[t][t][0.7]{$M_1$}
\psfrag{M2}[t][t][0.7]{$M_2$}
\psfrag{M3}[t][t][0.7]{$M_3$}
\psfrag{M4}[t][t][0.7]{$M_4$}
\psfrag{g0}[lb][lb][0.8]{$G_0(\omega)$}
\psfrag{g1}[br][br][0.8]{$G_1(\omega)$}
\psfrag{g2}[b][b][0.8]{$G_2(\omega)$}
\psfrag{g3}[bl][bl][0.8]{$G_3(\omega)$}
\psfrag{g4}[b][b][0.8]{$G_4(\omega)$}
 \psfrag{f0}[b][b][0.8]{$\varphi_0$}
 \psfrag{f1}[b][b][0.8]{$\varphi_1$}
 \psfrag{f2}[b][b][0.8]{$\varphi_2$}
 \psfrag{f3}[b][b][0.8]{$\varphi_3$}
 \psfrag{f4}[b][b][0.8]{$\varphi_4$}
 \psfrag{2o.}[b][b][0.6]{$\circ\circ$}
 \psfrag{2x.}[b][b][0.5]{$\times\!\times$}
 \psfrag{X.}[b][b][0.5]{$\times_i$}
\psfrag{a1}[tr][t][0.7]{$\omega_1^a$}
\psfrag{b1}[tl][t][0.7]{$\omega_1^b$}
\psfrag{a2}[t][t][0.7]{$\omega_2^a$}
\psfrag{b2}[tr][t][0.7]{$\omega_2^b$}
\psfrag{a3}[tl][t][0.7]{$\omega_3^a$}
\psfrag{b3}[t][t][0.7]{$\omega_3^b$}
\psfrag{a4}[tl][t][0.7]{$\omega_4^a$}
\psfrag{b4}[t][t][0.7]{$\omega_4^b$}
\psfrag{db}[][][0.7]{[db]}
\psfrag{Deg}[][][0.7]{[Deg]}
 \centering
 \includegraphics[clip,width=\columnwidth]{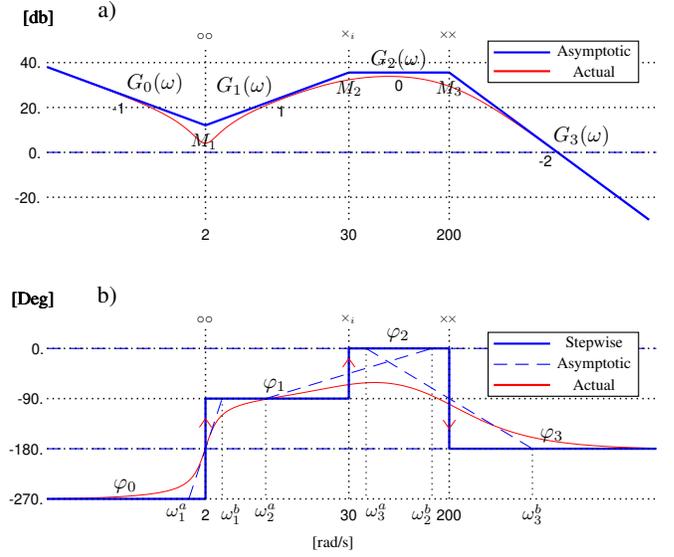}
   \setlength{\unitlength}{5.0mm}
 \psset{unit=\unitlength}
 \rput(-6,14.4){\small a)}
 \rput(-6,6.8){\small b)}
  \rput(0,0.25){\notsotiny [rad/s]}
  \rput(8.5,2.05){
  \psline[linewidth=0.42pt,fillcolor=white,fillstyle=solid](-4.5,10.45)(-0.5,10.45)(-0.5,11.55)(-4.5,11.55)(-4.5,10.45)
  \psline[linecolor=blue](-4.35,11.25)(-2.85,11.25)
  \rput(-1.68,11.25){\notsotiny Asymptotic}
  \psline[linecolor=red](-4.35,10.75)(-2.85,10.75)
  \rput(-1.68,10.75){\notsotiny Actual}
  }
  \rput(8.5,-6){
  \psline[linewidth=0.42pt,fillcolor=white,fillstyle=solid](-4.5,10.15)(-0.5,10.15)(-0.5,11.85)(-4.5,11.85)(-4.5,10.15)
  \psline[linecolor=blue](-4.35,11.55)(-2.85,11.55)
  \rput(-1.68,11.55){\notsotiny Stepwise}
  \psline[linewidth=0.42pt,linecolor=blue,linestyle=dashed](-4.35,11)(-2.85,11)
  \rput(-1.68,11){\notsotiny Asymptotic}
  \psline[linecolor=red](-4.35,10.45)(-2.85,10.45)
  \rput(-1.68,10.45){\notsotiny Actual}
  }
\vspace{-2mm}
   \caption{Asymptotic Bode plots of function $G(s)$ in \eqref{Gs_example} obtained using the direct method. (a) Asymptotic magnitude plot and actual magnitude plot. (b) Stepwise phase plot, asymptotic phase plot and actual phase plot. The critical frequencies are $\omega_1\!=\!2$ rad/s, $\omega_2\!=\!30$ rad/s, and $\omega_3\!=\!200$ rad/s, while the corresponding critical frequencies for the construction of the asymptotic phase plot are $\omega^a_1$, $\omega^b_1$, $\omega^a_2$, $\omega^b_2$, $\omega^a_3$, and $\omega^b_3$.}\label{BO_gs_2024_Example_1_Bode_Asymptotic_Plot}
\vspace{-2mm}
     \end{figure}

The critical gains $\M_k$ can either be computed from the approximating functions $G_k(s)$ in \eqref{Approx_Gs_example}
using \eqref{beta_K_example}:
  \[
 \begin{array}{r@{\;}c@{\;}l}
 \M_1
 &=&
\left| G_0(j2)\right|
= \left| G_1(j2)\right|
= \left| -2\,(j2))\right|
= 4 = 12 \db,
\\[1mm]
 \M_2
 &=&
\left| G_1(j30)\right|
=\left| G_2(j30)\right|
=\left| 60\right|
=60 \simeq 36 \db,
\\[1mm]
 \M_3
 &=&
\left| G_2(j200)\right|
= \left| G_3(j200)\right|
= \left| \frac{2.4\,10^6}{(j200)^2}\right|
= 60\simeq 36 \db,
\end{array}
\]
 or by using the recursive formula \eqref{beta_K_example_bis}:
  \[
 \begin{array}{r@{\;}c@{\;}l}
 \M_1
 &=&
 \left| G_0(s)\right|_{s=j2}
 =  \left| -\frac{8}{j2}\right|
 = 4,
\\[0mm]
 \M_2
 &=&
 \M_{1}\left(\frac{\omega_{1}}{\omega_{2}}\right)^{t_1\,}
 = 4\left(\frac{2}{30}\right)^{-1}
 =60,
\\[0mm]
 \M_3
 &=&
 \M_{2}\left(\frac{\omega_{2}}{\omega_{3}}\right)^{t_2\,}
 =  60\left(\frac{30}{200}\right)^{0}
 = 60,
\end{array}
\]
 where $t_1=-1$ and $t_2=0$ are the relative degrees of functions $G_1(s)$ and $G_2(s)$ in \eqref{Approx_Gs_example}, respectively.
 The segments $G_0(\omega)$, $G_1(\omega)$, $G_2(\omega)$ and $G_3(\omega)$ of the asymptotic Bode magnitude plot constructed using Property~\ref{bode_magn_prop}, which are shown in blue in Figure~\ref{BO_gs_2024_Example_1_Bode_Asymptotic_Plot}(a), approximate very well the real Bode magnitude plot $|G(j\omega)|$, shown in red in the  same figure.
The short notations $1$, $2$, $-1$ and $-2$ in Figure~\ref{BO_gs_2024_Example_1_Bode_Asymptotic_Plot}(a) denote the slopes $+20$ db/dec, $+40$ db/dec, $-20$ db/dec and $-40$ db/dec, respectively.
The frequency band phases $\varphi_k$ can be computed
using the recursive formula \eqref{varphi_k}:
\[ \begin{array}{@{}r@{\,}c@{\,}l@{}}
 \varphi_0
 &=&
 \arg(G_{0}(j\omega))
= \arg(-\frac{8}{j\omega})
=-\frac{3\pi}{2},
\\[2mm]
 \varphi_1
 &=&
\varphi_{0} \!+\!\Delta\varphi_1
 = \!-\!\frac{3\pi}{2} \!+\!
 \underbrace{\!r_1\!}_{1}
 \underbrace{\!S_1^{zp}\!}_{1}
 \underbrace{\!S_1^{st}\!}_{1}
 \underbrace{\!n^o_1\!}_{2}
\frac{\pi}{2}
 = -\frac{\pi}{2},
\\[2mm]
 \varphi_2
 &=&
\varphi_{1} \!+\!\Delta\varphi_2
 = \!-\!\frac{\pi}{2} \!+\!
 \underbrace{\!r_2\!}_{1}
 \underbrace{\!S_2^{zp}\!}_{-1}
 \underbrace{\!S_2^{st}\!}_{-1}
 \underbrace{\!n^o_2\!}_{1}
\frac{\pi}{2}
 = 0,
\\[2mm]
 \varphi_3
 &=&
\varphi_{2} \!+\!\Delta\varphi_3
 = 0 \!+\!
 \underbrace{\!r_3\!}_{2}
 \underbrace{\!S_3^{zp}\!}_{-1}
 \underbrace{\!S_3^{st}\!}_{1}
 \underbrace{\!n^o_3\!}_{1}
\frac{\pi}{2}
 = - \pi.
 \end{array}
\]
By plotting  the $4$ segments $\varphi_0(\omega)$, $\varphi_1(\omega)$, $\varphi_2(\omega)$ and $\varphi_3(\omega)$ and by connecting their endpoints, as described in Property~\ref{critical_gain_defini_prop}, the  stepwise asymptotic phase plot is obtained, which is shown in blue in Figure~\ref{BO_gs_2024_Example_1_Bode_Asymptotic_Plot}(b).
The asymptotic phase plot can be obtained from the stepwise phase plot by replacing the discontinuities at $\omega \in \Omega_c$ with the asymptotic interpolation associated with the dynamic elements acting at the considered critical frequency $\omega_k$. This replacement can be done by adding the two associated critical frequencies $\omega^a_k$ and $\omega^b_k$~\cite{Nuovo_9} in the vicinity of the critical frequency $\omega_k$:
\begin{equation}\label{wak_eq}
\begin{array}{l}
\omega^a_k
=\left\{\begin{array}{@{}l@{\hspace{2mm} \mbox{if} \hspace{2mm}}l}
\frac{\omega_k}{4.81} & \mbox{$p_i(s)$ is a first-order term}, \\[1mm]
\frac{\omega_k}{4.81^\delta} & \mbox{$p_i(s)$ is a second-order term},
\end{array}\right.
\\[4mm]
\omega^b_k
=\left\{\begin{array}{@{}l@{\hspace{2mm} \mbox{if} \hspace{2mm}}l}
\ts\omega_k 4.81 & \mbox{$p_i(s)$ is a first-order term}, \\[1mm]
\ts\omega_k 4.81^\delta & \mbox{$p_i(s)$ is a second-order term},
\end{array}\right.
\end{array}
\end{equation}
where $\delta$ is the damping coefficient of the second-order term.
The critical frequencies $\omega^a_k$ and $\omega^b_k$ associated with each critical frequency $\omega_k$, for $k \in \{1,\,2,\,3\}$, are shown in Figure~\ref{BO_gs_2024_Example_1_Bode_Asymptotic_Plot} in blue dotted lines.

\subsection{Second Case Study}\label{second_case_study_sect}

\begin{figure}[t]
 \psfrag{Phioo}[l][l][0.6]{$\varphi_\infty$}
 \psfrag{Phi0}[lb][lb][0.6]{$\varphi_0$}
 \psfrag{G0s}[l][l][0.6]{$G_0(s)$}
 \psfrag{Goos}[l][l][0.6]{$G_\infty(s)$}
 \psfrag{be}[lb][lb][0.5]{$M_1$}
 \psfrag{ga}[l][l][0.5]{$M_3$}
\psfrag{M0}[t][t][0.7]{}
\psfrag{M1}[t][t][0.7]{$M_1$}
\psfrag{M2}[t][t][0.7]{$M_2$}
\psfrag{M3}[t][t][0.7]{$M_3$}
\psfrag{M4}[t][t][0.7]{$M_4$}
\psfrag{g0}[lb][lb][0.8]{$G_0(\omega)$}
\psfrag{g1}[br][br][0.8]{$G_1(\omega)$}
\psfrag{g2}[b][b][0.8]{$G_2(\omega)$}
\psfrag{g3}[bl][bl][0.8]{$G_3(\omega)$}
\psfrag{g4}[b][b][0.8]{$G_4(\omega)$}
 \psfrag{f0}[b][b][0.8]{$\varphi_0$}
 \psfrag{f1}[b][b][0.8]{$\varphi_1$}
 \psfrag{f2}[b][b][0.8]{$\varphi_2$}
 \psfrag{f3}[b][b][0.8]{$\varphi_3$}
 \psfrag{f4}[b][b][0.8]{$\varphi_4$}
 \psfrag{2o.}[b][b][0.6]{$\circ\circ$}
 \psfrag{2x.}[b][b][0.5]{$\times\!\times$}
 \psfrag{X.}[b][b][0.5]{$\times_i$}
\psfrag{a1}[t][t][0.7]{$\omega_1^a$}
\psfrag{b1}[tl][t][0.7]{$\omega_1^b$}
\psfrag{a2}[rt][t][0.7]{$\omega_2^a$}
\psfrag{b2}[lb][][0.7]{$\omega_2^b$}
\psfrag{a3}[tr][t][0.7]{$\omega_3^a$}
\psfrag{b3}[tl][][0.7]{$\omega_3^b$}
\psfrag{a4}[tl][t][0.7]{$\omega_4^a$}
\psfrag{b4}[t][t][0.7]{$\omega_4^b$}
\psfrag{[db]}[][][0.7]{[db]}
\psfrag{[Deg]}[][][0.7]{[Deg]}
 \centering
 \includegraphics[clip,width=\columnwidth]{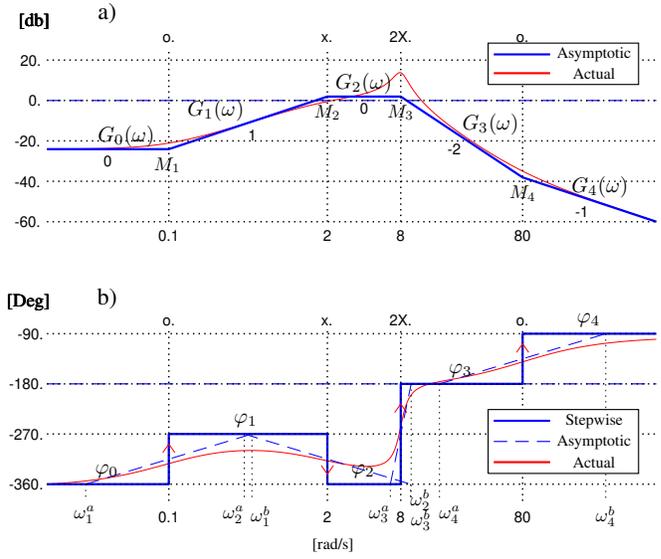}
   \setlength{\unitlength}{5.0mm}
 \psset{unit=\unitlength}
 \rput(-6,14.4){\small a)}
 \rput(-6,6.8){\small b)}
  \rput(0,0.25){\notsotiny [rad/s]}
  \rput(8.5,2.05){
  \psline[linewidth=0.42pt,fillcolor=white,fillstyle=solid](-4.5,10.45)(-0.5,10.45)(-0.5,11.55)(-4.5,11.55)(-4.5,10.45)
  \psline[linecolor=blue](-4.35,11.25)(-2.85,11.25)
  \rput(-1.68,11.25){\notsotiny Asymptotic}
  \psline[linecolor=red](-4.35,10.75)(-2.85,10.75)
  \rput(-1.68,10.75){\notsotiny Actual}
  }
  \rput(8.5,-8){
  \psline[linewidth=0.42pt,fillcolor=white,fillstyle=solid](-4.5,10.15)(-0.5,10.15)(-0.5,11.85)(-4.5,11.85)(-4.5,10.15)
  \psline[linecolor=blue](-4.35,11.55)(-2.85,11.55)
  \rput(-1.68,11.55){\notsotiny Stepwise}
  \psline[linewidth=0.42pt,linecolor=blue,linestyle=dashed](-4.35,11)(-2.85,11)
  \rput(-1.68,11){\notsotiny Asymptotic}
  \psline[linecolor=red](-4.35,10.45)(-2.85,10.45)
  \rput(-1.68,10.45){\notsotiny Actual}
  }
\vspace{-2mm}
   \caption{
Asymptotic Bode plots of function $G(s)$ in \eqref{Gs_example_2} obtained using the direct method. (a) Asymptotic magnitude plot and actual magnitude plot. (b) Stepwise phase plot, asymptotic phase plot and actual phase plot. The critical frequencies are $\omega_1=0.1$ rad/s, $\omega_2=2$ rad/s, $\omega_3=8$ rad/s, and $\omega_4=80$ rad/s, while the corresponding critical frequencies for the construction of the asymptotic phase plot are $\omega^a_1$, $\omega^b_1$, $\omega^a_2$, $\omega^b_2$, $\omega^a_3$, $\omega^b_3$, $\omega^a_4$, and $\omega^b_4$.
\label{BO_gs_2024_18_06_2026_Bode_Asymptotic_Plot_bis}
}
\vspace{-2mm}
\end{figure}

Reference is made to the following transfer function:
\begin{equation}\label{Gs_example_2}
 G(s)=
\frac{(s+0.1)(s+80)}{(s+2)(s^2-2s+64)}.
\end{equation}
 In this case, the polynomial terms are: $p_1(s)=(s+0.1)$,  $p_2(s)=(s+80)$,  $p_3(s)=(s+2)$ and  $p_4(s)=(s^2-2s+64)$.
  The set $\Omega_c$ of critical frequencies
  is  $\Omega_c=\{0.1,\;2,\;8,\;80\}$.
By applying \eqref{G_k_s} and using \eqref{Gs_example_2}, the following approximating functions $G_k(s)$ are obtained:
\[
 \begin{array}{@{}r@{\,}c@{\,}l@{\;\;\mbox{for}\;\;}c@{}}
 G_0(s)
 &=&
 \frac{({\color{verylightgrey}\cancel{s}}+0.1)({\color{verylightgrey}\cancel{s}}+80)}{({\color{verylightgrey}\cancel{s}}+2)({\color{verylightgrey}\cancel{s^2}-\cancel{2s}}+64)}
 =\frac{1}{16}
 & 
 \omega \leq 0.1,
\\[1mm]
 G_1(s)
 &=&
 \frac{(s{+\color{verylightgrey}\cancel{0.1}})({\color{verylightgrey}\cancel{s}}+80)}{({\color{verylightgrey}\cancel{s}}+2)({\color{verylightgrey}\cancel{s^2}-\cancel{2s}}+64)}
 =\frac{5\,s}{8}
  &
 0.1\leq \omega \leq 2,
\\[1mm]
 G_2(s)
 &=&
 \frac{(s{+\color{verylightgrey}\cancel{0.1}})({\color{verylightgrey}\cancel{s}}+80)}{(s{\color{verylightgrey}+\cancel{2}})({\color{verylightgrey}\cancel{s^2}-\cancel{2s}}+64)}
 =\frac{5}{4}
  &
 2\leq \omega \leq 8,
 \\[1mm]
 G_3(s)
 &=&
 \frac{(s{\color{verylightgrey}+\cancel{0.1}})({\color{verylightgrey}\cancel{s}}+80)}{(s{\color{verylightgrey}+\cancel{2}})(s^2{\color{verylightgrey}-\cancel{2s}+\cancel{64}})}
 =\frac{80}{s^2}
  &
 8\leq \omega \leq 80,
\\[1mm]
 G_4(s)
 &=&
 \frac{(s{\color{verylightgrey}+\cancel{0.1}})(s{\color{verylightgrey}+\cancel{80}})}{(s{\color{verylightgrey}+\cancel{2}})(s^2{\color{verylightgrey}-\cancel{2s}+\cancel{64}})}
 =\frac{1}{s}
  &
 \omega \geq 80.
 \end{array}
 \]
The critical gains $\M_k$ can either be computed from the approximating functions $G_k(s)$ in \eqref{Gs_example_2}
using \eqref{beta_K_example}:
  \[
 \begin{array}{r@{\;}c@{\;}l}
 \M_1
 &=&
\left| G_0(j 0.1)\right|
= \frac{1}{16} = -24.08 \db,
\\[1mm]
 \M_2
 &=&
\left| G_1(j 2)\right|
=\left| \frac{5\,j2}{8}\right|
=\frac{5}{4} \simeq 1.93 \db,
\\[1mm]
 \M_3
 &=&
\left| G_2(j 8)\right|
= \left| \frac{5}{4}\right|
=\frac{5}{4} \simeq 1.93 \db,
\\[1mm]
 \M_4
 &=&
\left| G_3(j 80)\right|
= \left| \frac{80}{(j80)^2}\right|
= \frac{1}{80}\simeq -38.06 \db,
\end{array}
\]
 or by using the recursive formula \eqref{beta_K_example_bis}:
  \[
 \begin{array}{r@{\;}c@{\;}l}
 \M_1
 &=&
\left| G_0(j 0.1)\right|
= \frac{1}{16},
\\[0mm]
 \M_2
 &=&
 \M_{1}\left(\frac{\omega_{1}}{\omega_{2}}\right)^{t_1\,}
 = \frac{1}{16}\left(\frac{0.1}{2}\right)^{-1}
 =\frac{5}{4},
\\[0mm]
 \M_3
 &=&
 \M_{2}\left(\frac{\omega_{2}}{\omega_{3}}\right)^{t_2\,}
 =  \frac{5}{4}\left(\frac{2}{8}\right)^{0}
 = \frac{5}{4},
\\[0mm]
 \M_4
 &=&
 \M_{3}\left(\frac{\omega_{3}}{\omega_{4}}\right)^{t_3\,}
 =  \frac{5}{4}\left(\frac{8}{80}\right)^{2}
 = \frac{1}{80}.
\end{array}
\]
By plotting the 5 segments $G_0(\omega)$, $G_1(\omega)$, $G_2(\omega)$,  $G_3(\omega)$ and $G_4(\omega)$ constructed using Property~\ref{bode_magn_prop}, the asymptotic Bode magnitude plot shown in blue in Figure~\ref{BO_gs_2024_18_06_2026_Bode_Asymptotic_Plot_bis}(a) is obtained.
The frequency band phases $\varphi_k$ of function $G(s)$ computed
using the recursive formula \eqref{varphi_k} are the following:
\[ \begin{array}{@{}r@{\,}c@{\,}l@{}}
 \varphi_0
 &=&
 \arg(G_{0}(j\omega))
= \arg(\frac{1}{16})
= 0 \equiv - 2\pi,
\\[3mm]
 \varphi_1
 &=&
\varphi_{0} \!+\!\Delta\varphi_1
 = \!-\!2\pi \!+\!
 \underbrace{\!r_1\!}_{1}
 \underbrace{\!S_1^{zp}\!}_{1}
 \underbrace{\!S_1^{st}\!}_{1}
 \underbrace{\!n^o_1\!}_{1}
\frac{\pi}{2}
 = -\frac{3\pi}{2},
 \end{array}
\]
\vspace{-4.8mm}
\[
\begin{array}{@{}r@{\,}c@{\,}l@{}}
 \varphi_2
 &=&
\varphi_{1} \!+\!\Delta\varphi_3
 = \!-\!\frac{3\pi}{2} \!+\!
 \underbrace{\!r_3\!}_{1}
 \underbrace{\!S_3^{zp}\!}_{-1}
 \underbrace{\!S_3^{st}\!}_{1}
 \underbrace{\!n^o_3\!}_{1}
\frac{\pi}{2}
 = -2\pi,\\[2mm]
 \varphi_3
 &=&
\varphi_{2} \!+\!\Delta\varphi_4
 = -2\pi \!+\!
 \underbrace{\!r_4\!}_{1}
 \underbrace{\!S_4^{zp}\!}_{-1}
 \underbrace{\!S_4^{st}\!}_{-1}
 \underbrace{\!n^o_4\!}_{2}
\frac{\pi}{2}
 = - \pi,
\\[2mm]
 \varphi_4
 &=&
\varphi_{3} \!+\!\Delta\varphi_2
 = - \pi \!+\!
 \underbrace{\!r_2\!}_{1}
 \underbrace{\!S_2^{zp}\!}_{1}
 \underbrace{\!S_2^{st}\!}_{1}
 \underbrace{\!n^o_2\!}_{1}
\frac{\pi}{2}
 = - \frac{\pi}{2}.
 \end{array}
\]
By plotting the $5$ segments $\varphi_0(\omega)$, $\varphi_1(\omega)$, $\varphi_2(\omega)$, $\varphi_3(\omega)$ and $\varphi_4(\omega)$ and by connecting their endpoints, as described in Property~\ref{critical_gain_defini_prop}, the stepwise asymptotic phase plot shown in blue in Figure~\ref{BO_gs_2024_18_06_2026_Bode_Asymptotic_Plot_bis}(b) is obtained.
The critical frequencies $\omega^a_k$ and $\omega^b_k$ associated with each critical frequency $\omega_k$
can be computed as in \eqref{wak_eq}, and are shown
in Figure~\ref{BO_gs_2024_18_06_2026_Bode_Asymptotic_Plot_bis}.
The actual bode magnitude and phase plots are shown in red 
in Figure~\ref{BO_gs_2024_18_06_2026_Bode_Asymptotic_Plot_bis}(a) and Figure~\ref{BO_gs_2024_18_06_2026_Bode_Asymptotic_Plot_bis}(b), respectively.

\section{Comparison with the Classical
Method}\label{comparison_sect}
In this section, the comparison of the new direct method for constructing asymptotic Bode plots described in Sec.~\ref{Asymptotic_Bode_Plots} against the standard method \cite{Nuovo_6_bis}, consisting in the summation of the contributions coming from the single factors, is addressed with reference to the following case study function:
\begin{equation}\label{example_2}
G(s)=\frac{10(s-1)}{s(s+1)(s^{2}+8s+25)}.
\end{equation}
\subsection{Direct Method}\label{Direct_Method_sect}

 The polynomial terms of function $G(s)$ are: $p_1(s)=(s-1)$,  $p_2(s)=(s+1)$ and  $p_3(s)=(s^{2}+8s+25)$, while
 the set $\Omega_c$ of critical frequencies is  $\Omega_c=\{1,\;5\}$.
By applying \eqref{G_k_s} to \eqref{example_2}, the approximating functions $G_k(s)$ are the following:
%
  \begin{equation}\label{Approx_Gs_example_2}
 \begin{array}{@{}r@{\,}c@{\,}l@{\;\;\mbox{for}\;\;}c@{}}
 G_0(s)
 &=&
 \frac{10({\color{verylightgrey}\cancel{s}}-1)}{s({\color{verylightgrey}\cancel{s}}+1)({\color{verylightgrey}\cancel{s^{2}}+\cancel{8s}}+25)}
 =-\frac{10}{25\,s}
 &  
 \omega \leq 1,
\\[1mm]
 G_1(s)
 &=&
 \frac{10(s{\color{verylightgrey}-\cancel{1}})}{s(s{\color{verylightgrey}+\cancel{1}})({\color{verylightgrey}\cancel{s^{2}}+\cancel{8s}}+25)}
 =\frac{10}{25\,s}
  &
 1\leq \omega \leq 5,
\\[1mm]
 G_2(s)
 &=&
 \frac{10(s{\color{verylightgrey}-\cancel{1}})}{s(s{\color{verylightgrey}+\cancel{1}})(s^{2}{\color{verylightgrey}+\cancel{8s}+\cancel{25}})}
 =\frac{10}{s^3}
  &
 \omega \geq 5.
 \end{array}
\end{equation}
The critical gains $\M_k$
using \eqref{beta_K_example} are the following:
  \[
  \begin{array}{c}
 \M_1
 \!=\!
 \left| G_0(s)\right|_{s=j1}
 =\frac{10}{25},
\hspace{8mm}
 \M_2
 \!=\!
 \left| G_1(s)\right|_{s=j5}
 =\frac{2}{25}.
\end{array}
\]
\begin{figure}[t]
\begin{center}
 \psfrag{Phioo}[l][l][0.6]{$\varphi_\infty$}
 \psfrag{Phi0}[lb][lb][0.6]{$\varphi_0$}
 \psfrag{G0s}[l][l][0.6]{$G_0(s)$}
 \psfrag{Goos}[l][l][0.6]{$G_\infty(s)$}
 \psfrag{be}[lb][lb][0.5]{$M_1$}
 \psfrag{ga}[l][l][0.5]{$M_3$}
\psfrag{M0}[t][t][0.7]{}
\psfrag{M1}[t][t][0.7]{$M_1$}
\psfrag{M2}[t][t][0.7]{$M_2$}
\psfrag{M3}[t][t][0.7]{$M_3$}
\psfrag{M4}[t][t][0.7]{$M_4$}
\psfrag{g0}[lb][lb][0.8]{$G_0(\omega)$}
\psfrag{g1}[br][br][0.8]{$G_1(\omega)$}
\psfrag{g2}[b][b][0.8]{$G_2(\omega)$}
\psfrag{g3}[bl][bl][0.8]{$G_3(\omega)$}
\psfrag{g4}[b][b][0.8]{$G_4(\omega)$}
 \psfrag{f0}[b][b][0.8]{$\varphi_0$}
 \psfrag{f1}[b][b][0.8]{$\varphi_1$}
 \psfrag{f2}[b][b][0.8]{$\varphi_2$}
 \psfrag{f3}[b][b][0.8]{$\varphi_3$}
 \psfrag{f4}[b][b][0.8]{$\varphi_4$}
 \psfrag{2o.}[b][b][0.6]{$\circ\circ$}
 \psfrag{2x.}[b][b][0.5]{$\times\!\times$}
 \psfrag{X.}[b][b][0.5]{$\times_i$}
\psfrag{a1}[t][t][0.7]{$\omega_1^a$}
\psfrag{b1}[tr][t][0.7]{$\omega_1^b\;$}
\psfrag{a2}[t][t][0.7]{$\omega_2^a$}
\psfrag{b2}[t][t][0.7]{$\omega_2^b$}
\psfrag{a3}[tr][t][0.7]{$\omega_3^a$}
\psfrag{b3}[tl][t][0.7]{$\omega_3^b$}
\psfrag{a4}[tl][t][0.7]{$\omega_4^a$}
\psfrag{b4}[t][t][0.7]{$\omega_4^b$}
\psfrag{db}[][][0.7]{{[db]}}
\psfrag{Deg}[][][0.7]{[Deg]}
\psfrag{2o.}[b][b][0.6]{$\circ\circ$}
 \psfrag{2x.}[b][b][0.5]{$\times\!\times$}
 \psfrag{xO.}[b][b][0.5]{$\times \circ_i$}
 \psfrag{be}[r][r][0.7]{$\delta$}
 \psfrag{G0s}[b][b][0.7]{$\left|G_0(j\omega)\right|$}
 \psfrag{Goos}[b][b][0.7]{$\left|G_\infty(j\omega)\right|$}
 \psfrag{ga}[b][b][0.7]{$\gamma=\M$}
 \psfrag{Phi0}[b][b][0.7]{$\varphi_0$}
 \psfrag{Phioo}[b][b][0.7]{$\varphi_\infty$}
 \includegraphics[clip,width=\columnwidth]{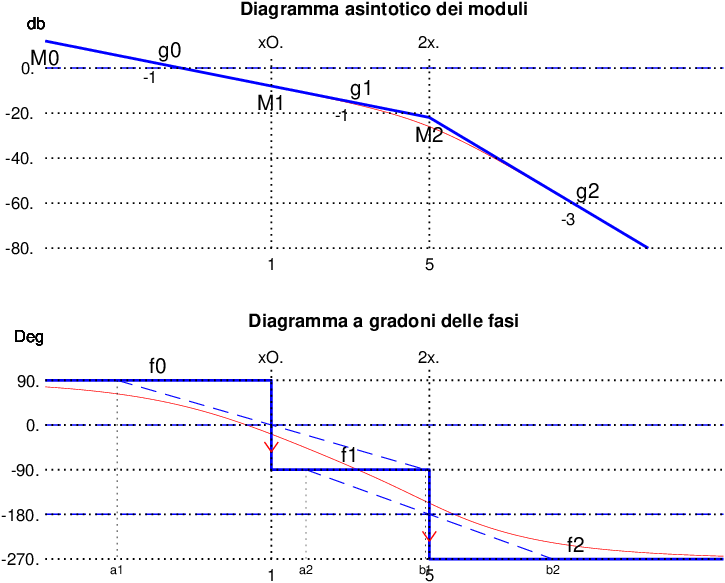}
   \setlength{\unitlength}{5.0mm}
 \psset{unit=\unitlength}
 \rput(-6,14.4){\small a)}
 \rput(-6,6.8){\small b)}
  \rput(0,0.25){\notsotiny [rad/s]}
  \rput(8.5,2.05){
  \psline[linewidth=0.42pt,fillcolor=white,fillstyle=solid](-4.5,10.45)(-0.5,10.45)(-0.5,11.55)(-4.5,11.55)(-4.5,10.45)
  \psline[linecolor=blue](-4.35,11.25)(-2.85,11.25)
  \rput(-1.68,11.25){\notsotiny Asymptotic}
  \psline[linecolor=red](-4.35,10.75)(-2.85,10.75)
  \rput(-1.68,10.75){\notsotiny Actual}
  }
  \rput(8.5,-6){
  \psline[linewidth=0.42pt,fillcolor=white,fillstyle=solid](-4.5,10.15)(-0.5,10.15)(-0.5,11.85)(-4.5,11.85)(-4.5,10.15)
  \psline[linecolor=blue](-4.35,11.55)(-2.85,11.55)
  \rput(-1.68,11.55){\notsotiny Stepwise}
  \psline[linewidth=0.42pt,linecolor=blue,linestyle=dashed](-4.35,11)(-2.85,11)
  \rput(-1.68,11){\notsotiny Asymptotic}
  \psline[linecolor=red](-4.35,10.45)(-2.85,10.45)
  \rput(-1.68,10.45){\notsotiny Actual}
  }
\end{center}
\vspace{-2mm}
    \caption{Asymptotic Bode plots of function $G(s)$ in \eqref{example_2} obtained using the direct method. (a) Asymptotic magnitude plot and actual magnitude plot. (b) Stepwise phase plot, asymptotic phase plot and actual phase plot. The critical frequencies are $\omega_1=1$ rad/s and $\omega_2=5$ rad/s, while the corresponding critical frequencies for the construction of the asymptotic phase plot are $\omega^a_1$, $\omega^b_1$, $\omega^a_2$ and $\omega^b_2$.
    }
\label{BO_gs_2024_Example_A_B_Bode_Asymptotic_Plot}
\vspace{-2mm}
 \end{figure}
By plotting the segments $G_0(\omega)$, $G_1(\omega)$ and $G_2(\omega)$ constructed using Property~\ref{bode_magn_prop}, the asymptotic Bode magnitude plot shown
in Figure~\ref{BO_gs_2024_Example_A_B_Bode_Asymptotic_Plot}(a) is obtained.
In this case, the frequency band phases $\varphi_k$ of function $G(s)$ computed  using
\eqref{varphi_k} in Property~\ref{critical_gain_defini_prop} are the following:
\begin{equation}\label{varphi_k_rec_form}
 \begin{array}{@{}r@{\,}c@{\,}l@{}}
 \varphi_0
 &=&
 \arg(G_{0}(j\omega))
= \arg(-\frac{10}{25\,s})
= -\frac{3\pi}{2},
\\[1mm]
 \varphi_1
 &=&
\varphi_{0} + \Delta\varphi_1 + \Delta\varphi_2
\\[1mm]
 &=&
 -\frac{3\pi}{2}  +
 \underbrace{\!r_1\!}_{1}
 \underbrace{\!S_1^{zp}\!}_{1}
 \underbrace{\!S_1^{st}\!}_{-1}
 \underbrace{\!n^o_1\!}_{1}
\frac{\pi}{2} + \ldots
\\[1mm]
 &&
 \hspace{-2mm}
\ldots
 +
 \underbrace{\!r_2\!}_{1}
 \underbrace{\!S_2^{zp}\!}_{-1}
 \underbrace{\!S_2^{st}\!}_{1}
 \underbrace{\!n^o_2\!}_{1}
\frac{\pi}{2}
 = -\frac{5\pi}{2},
\\[2mm]
 \varphi_2
 &=&
  \varphi_{1} \!+\!\Delta\varphi_3
 = -\frac{5\pi}{2} \!+\!
 \underbrace{\!r_3\!}_{1}
 \underbrace{\!S_3^{zp}\!}_{-1}
 \underbrace{\!S_3^{st}\!}_{1}
 \underbrace{\!n^o_3\!}_{2}
\frac{\pi}{2}
 = -\frac{7\pi}{2}.
 \end{array}
\end{equation}
By plotting the segments $\varphi_0(\omega)$, $\varphi_1(\omega)$ and $\varphi_2(\omega)$ and by connecting their endpoints, as described in Property~\ref{critical_gain_defini_prop}, the stepwise asymptotic phase plot shown
in Figure~\ref{BO_gs_2024_Example_A_B_Bode_Asymptotic_Plot}(b) is obtained.
The critical frequencies $\omega^a_k$ and $\omega^b_k$ associated with each critical frequency $\omega_k$
can be computed as in \eqref{wak_eq},  and are shown
in Figure~\ref{BO_gs_2024_Example_A_B_Bode_Asymptotic_Plot}.

\subsection{Standard Method}\label{Classical_Method_sect}
The considered case study in \eqref{example_2} is composed of the following different components:
\begin{equation}\label{components_eq}
\begin{array}{c}
 { K}= -\frac{10}{25},
 \hspace{0.4cm}
 { G_{1}}(s)=(1-s),
 \hspace{0.4cm}
 { G_{2}}(s)=\frac{1}{s},
\\[2mm]
 { G_{3}}(s)=\frac{1}{(1+s)},
 \hspace{0.4cm}
 { G_{4}}(s)=\frac{1}{(1+\frac{8s}{25}+\frac{s^{2}}{25})}.
\end{array}
\end{equation}
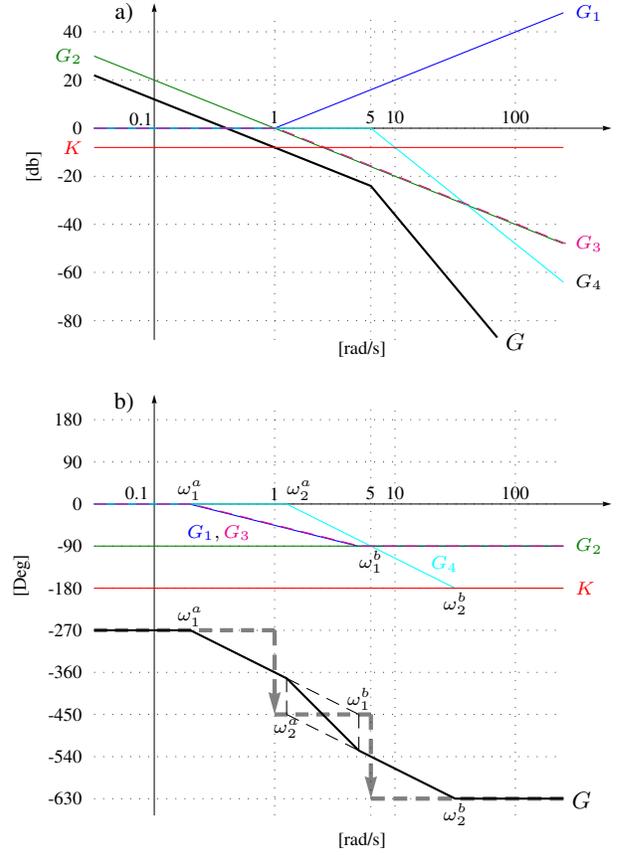
\begin{figure}[t]
\begin{center}
  \setlength{\unitlength}{1.6mm}
 \SpecialCoor
 \psset{xunit=\unitlength}
 \psset{yunit=0.8\unitlength}
 \psset{arrowlength=1.8}
 \psset{arrowinset=.1}
 \begin{picture}(48,32)(1.6,-17.25)
 { \scriptsize
 \rput(7.5,12){\small a)}
 \rput(40,-22.25){\small $G$}
 \rput{90}(0,-4){[db]}
 \rput(27.25,-23){[rad/s]}
 \psline[linewidth=0.2pt]{->}(5,0)(48,0)
 \psline[linewidth=0.2pt]{->}(10,-22)(10,13)
 \multirput(5,-20)(0,5){7}{\psline[linewidth=0.4pt,linestyle=dotted](39,0)}
 \multirput(20,-22)(10,0){3}{\psline[linewidth=0.4pt,linestyle=dotted](0,34)}
 \psline[linewidth=0.2pt,linestyle=dotted](28,-22)(28,12)
 \psline[linewidth=0.8pt](5,5.5)(28,-6)(38.5,-21.75)
 \psline[linewidth=0.4pt,linecolor=verde](5,7.5)(44,-12)
 \psline[linewidth=0.4pt,linecolor=blue](5,0)(20,0)(44, 12)
 \psline[linewidth=0.4pt,linecolor=arancione](5,0)(28,0)(44,-16)
 \psline[linewidth=0.4pt,linestyle=dashed,linecolor=magenta](5,0)(20.2,0)(44.2,-12)
 \psline[linewidth=0.4pt,linecolor=red](5,-2)(44,-2)
 \rput(4,-20){\makebox(0,0)[r]{-80}}
 \rput(4,-15){\makebox(0,0)[r]{-60}}
 \rput(4,-10){\makebox(0,0)[r]{-40}}
 \rput(4, -5){\makebox(0,0)[r]{-20}}
 \rput(4, -2){\makebox(0,0)[r]{\color{red}$K$}}
 \rput(4,  0){\makebox(0,0)[r]{  0}}
 \rput(4,  5){\makebox(0,0)[r]{ 20}}
 \rput(4, 10){\makebox(0,0)[r]{ 40}}
 \rput(45,12){\makebox(0,0)[l]{$\color{blue}G_{1}$}}
 \rput(4,7.5){\makebox(0,0)[r]{$\color{verde}G_{2}$}}
 \rput(45,-12){\makebox(0,0)[l]{$\color{magenta}G_{3}$}}
 \rput(45,-16){\makebox(0,0)[l]{$G_{4}$}}
 \rput( 9.85,0.25){\makebox(0,0)[br]{0.1}}
 \rput(20, 0.5){\makebox(0,0)[b]{1}}
 \rput(28,0.5){\makebox(0,0)[b]{5}}
 \rput(30,0.5){\makebox(0,0)[b]{10}}
 \rput(40,0.5){\makebox(0,0)[b]{100}}
 }
 \end{picture}
 \setlength{\unitlength}{1.6mm}
 \SpecialCoor
 \psset{xunit=\unitlength}
 \psset{yunit=0.7\unitlength}
 \psset{arrowlength=1.8}
 \psset{arrowinset=.1}
 \begin{picture}(48,42)(1.6,-28)
 { \scriptsize
 \rput(7.5,12){\small b)}
  \rput(45.5,-35.25){\small $G$}
 \rput{90}(-1,-8.5){[Deg]}
 \rput(27.25,-40){[rad/s]}
 \psline[linewidth=0.2pt]{->}(5,0)(48,0)
 \psline[linewidth=0.2pt]{->}(10,-37)(10,13)
 \multirput(5,-35)(0,5){10}{\psline[linewidth=0.4pt,linestyle=dotted](39,0)}
 \multirput(20,-37)(10,0){3}{\psline[linewidth=0.4pt,linestyle=dotted](0,49)}
 \psline[linewidth=1.6pt,linestyle=dashed,linecolor=gray](5,-15)(20,-15)
 \psline[linewidth=1.6pt,linestyle=dashed,linecolor=gray]{->}(20,-15)(20,-25)
 \psline[linewidth=1.6pt,linestyle=dashed,linecolor=gray](20,-25)(28,-25)
 \psline[linewidth=1.6pt,linestyle=dashed,linecolor=gray]{->}(28,-25)(28,-35)
 \psline[linewidth=1.6pt,linestyle=dashed,linecolor=gray](28,-35)(44,-35)
 \psline[linewidth=0.8pt](5,-15)(13,-15)(21,-20.71)(27,-29.29)(35,-35)(44,-35)
 \psline[linewidth=0.2pt,linestyle=dotted](28,-37)(28,12)
 \psline[linewidth=0.4pt,linecolor=blue](5,0)(12.9,0)(26.9,-5)(44,-5)
 \psline[linewidth=0.4pt,linecolor=arancione](5,0)(21,0)(35,-10)(44,-10)
 \psline[linewidth=0.4pt,linecolor=verde](5,-5)(44,-5)       
 \psline[linewidth=0.4pt,linestyle=dashed,linecolor=magenta](5,0)(13.1,0)(27.1,-5)(44,-5)
 \psline[linewidth=0.4pt,linecolor=red](5,-10)(44,-10)     
 \psline[linewidth=0.4pt,linestyle=dashed](21,-25)(35,-35)(44,-35)
   \psline[linewidth=0.4pt,linestyle=dashed](21,-20.71)(27,-25)
   \psline[linewidth=0.4pt,linestyle=dashed](21,-25)(21,-20.71)
   \psline[linewidth=0.4pt,linestyle=dashed](27,-25)(27,-29.29)
 \rput(4,-35){\makebox(0,0)[r]{-630}}
 \rput(4,-30){\makebox(0,0)[r]{-540}}
 \rput(4,-25){\makebox(0,0)[r]{-450}}
 \rput(4,-20){\makebox(0,0)[r]{-360}}
 \rput(4,-15){\makebox(0,0)[r]{-270}}
 \rput(4,-10){\makebox(0,0)[r]{-180}}
 \rput(4, -5){\makebox(0,0)[r]{-90}}
 \rput(4,  0){\makebox(0,0)[r]{  0}}
 \rput(4,  5){\makebox(0,0)[r]{90}}
 \rput(4, 10){\makebox(0,0)[r]{180}}
 \rput(18, -2.5){\makebox(0,0)[tr]{$\color{blue}G_{1}\color{black},\color{magenta}G_{3}$}}
 \rput(33, -8){\makebox(0,0)[bl]{$\color{arancione}G_{4}$}}
 \rput(45, -5){\makebox(0,0)[l]{\color{verde}$G_{2}$}}
 \rput(45,-10){\makebox(0,0)[l]{\color{red}$K$}}
 \rput(13,0.5){\makebox(0,0)[b]{$\omega^a_1$}}
 \rput(13,-14.5){\makebox(0,0)[b]{$\omega^a_1$}}
 \rput(27,-5.5){\makebox(0,0)[tl]{$\omega^b_1$}}
 \rput(27,-24.5){\makebox(0,0)[b]{$\omega^b_1$}}
 \rput(21,0.5){\makebox(0,0)[bl]{$\omega^a_2$}}
 \rput(21,-25.5){\makebox(0,0)[t]{$\omega^a_2$}}
 \rput(35,-10.5){\makebox(0,0)[t]{$\omega^b_2$}}
 \rput(35,-35.5){\makebox(0,0)[t]{$\omega^b_2$}}
 \rput( 9.5,0.5){\makebox(0,0)[br]{0.1}}
 \rput(20, 0.5){\makebox(0,0)[b]{1}}
 \rput(28,0.5){\makebox(0,0)[b]{5}}
 \rput(30,0.5){\makebox(0,0)[b]{10}}
 \rput(40,0.5){\makebox(0,0)[b]{100}}
 }
 \end{picture}
\end{center}
\vspace{-2.8mm}
\caption{Asymptotic Bode plots of function $G(s)$ in \eqref{example_2} obtained using the standard method of adding the contributions from the single components. (a) Asymptotic magnitude plot of the components \eqref{components_eq} of function $G(s)$ in \eqref{example_2} and asymptotic magnitude plot (black) of function $G(s)$ in \eqref{example_2}. (b) Asymptotic phase plot of the components \eqref{components_eq} of function $G(s)$ in \eqref{example_2}, stepwise (grey) and asymptotic (black) phase plots of function $G(s)$ in \eqref{example_2}. The critical frequencies are $\omega_1=1$ rad/s and $\omega_2=5$ rad/s, while the corresponding critical frequencies for the construction of the asymptotic phase plot are $\omega^a_1$, $\omega^b_1$, $\omega^a_2$ and $\omega^b_2$.}\label{magnitude_second}
\vspace{-2mm}
 \end{figure}
The asymptotic magnitude and phase Bode plots of the components $K$, $G_1(j\omega)$, $G_2(j\omega)$, $G_3(j\omega)$ and $G_4(j\omega)$ are shown in  Figure~\ref{magnitude_second}(a) and Figure~\ref{magnitude_second}(b), respectively.

As far as the construction of the asymptotic magnitude plot is concerned, the contributions from the different components $|K|$, $|G_1(j\omega)|$, $|G_2(j\omega)|$, $|G_3(j\omega)|$ and $|G_4(j\omega)|$ need to be added, since they are expressed in decibels. By carrying out this procedure within the whole frequency range, the resulting asymptotic magnitude bode plot of $G(j\omega)$ shown in black in Figure~\ref{magnitude_second}(a) is obtained.

As far as the construction of the asymptotic phase plot is concerned, the contributions from the different components  $\mbox{arg}(K)$, $\mbox{arg}(G_1(j\omega))$, $\mbox{arg}(G_2(j\omega))$, $\mbox{arg}(G_3(j\omega))$ and $\mbox{arg}(G_4(j\omega))$ need to be added. By carrying out this procedure within the whole frequency range, the resulting stepwise phase plot of $G(j\omega)$ shown in grey dashed line in Figure~\ref{magnitude_second}(b) is obtained. The asymptotic phase plot can then be derived by calculating the critical frequencies $\omega^a_k$ and $\omega^b_k$
as in \eqref{wak_eq}, and is shown in black solid line  in Figure~\ref{magnitude_second}(b).

\subsection{Comparative Discussion}\label{Comparative_Discussion_sect}
The following discussion can be made. 1) The considered standard method for the construction of asymptotic Bode plots is less systematic compared to the proposed direct method;  2) The considered standard method for the construction of asymptotic Bode plots requires more intermediate steps, and may be less precise and more time-consuming than the proposed direct method. The aforementioned statement 1) can be justified as follows: the standard method requires the initial detailed analysis of function $G(s)$ to identify its components, which vary from case to case, and the plotting of the asymptotic diagrams of each component. Conversely, the proposed direct method requires the calculation of the generalized approximating functions $G_k(s)$ and the systematic application of a few properties, with no need to analyze in detail the different factors or to plot their diagrams.
Furthermore, the standard method may be less precise and more subject to mistakes, especially if the number of components of function $G(s)$ increases. This is due to the intermediate step of plotting the asymptotic diagrams of each component of function $G(s)$, which then need to be added one-by-one to obtain the final asymptotic Bode plots of function $G(s)$. Indeed, the number of plots in Figure~\ref{magnitude_second} may be confusing, even if the number of components in the considered case study is relatively low. On the contrary, the proposed direct method requires neither the detailed analysis of the single components of $G(s)$, nor the plotting of their asymptotic Bode plots. In fact, the asymptotic Bode plots of function $G(s)$ can be obtained directly by drawing the sequences of segments as described in Property~\ref{bode_magn_prop} and Property~\ref{critical_gain_defini_prop} for the magnitude and the phase plots, respectively.

The aforementioned statements 1) and 2) also hold when comparing the proposed direct method against the second standard method \cite{Nuovo_6} described in the introduction.
Since, in the second standard method, the asymptotic magnitude plot is constructed by analyzing the amplitude variations introduced by the different factors, this method is less systematic than the proposed direct method.
Furthermore, the imprecision is likely to accumulate as the number of factors increases and the asymptotic magnitude plot construction goes on. On the contrary, since the proposed direct method identifies all the points where the slope in the asymptotic magnitude plot
changes at once at the beginning of the construction, it may be less time-consuming and more precise than the second standard method too.

\section{Conclusion}\label{Conclusion_sect}

This paper has addressed the proposal of a new direct method for the construction of asymptotic Bode plots, which are essential in control and systems engineering to perform an initial qualitative analysis of the considered systems. The proposed direct method is based on the use of generalized approximating functions, which enable to directly construct the asymptotic magnitude and phase plots of the complete transfer function and do not require the detailed analysis or the plots construction of each factor. The proposed direct method is more systematic, may be more precise and less-time consuming than standard methods, especially if the number of factors composing the transfer function increases.

\end{document}